\documentclass[aps,prl,showpacs,preprintnumbers,amsmath,amssymb,10pt]{article}
\usepackage{graphicx}
\usepackage{ifpdf}

\usepackage{caption}
\usepackage{color}
\title{Review on recent groundbreaking experiments on quantum communication with orthogonal states}
\author{A. Avella, G. Brida, D. Carpentras, A. Cavanna, I. P. Degiovanni, \\M. Genovese, M. Gramegna and P. Traina.\\ \\
INRIM, Strada delle cacce 91, I-10135 Turin, Italy.}
\newcommand{\ket}[1]{\mbox{\ensuremath{|#1\rangle}}}

\newcommand{\bea}{\vspace{0.25cm}\begin{eqnarray}}
\newcommand{\be}{\begin{equation}}
\newcommand{\ee}{\end{equation}}
\newcommand{\eea}{\end{eqnarray}}
\def\PRL{{Phys. Rev. Lett.} }
\def\PRA{{Phys. Rev.} A }

\begin{document}
\maketitle
\begin{abstract}
In recent years Quantum Key Distribution (QKD) has emerged as the most paradigmatic example of
Quantum technology allowing
the realization of intrinsically secure communication links over
hundreds of kilometers. Beyond its commercial interest QKD also has high conceptual relevance
 in the study of quantum information theory and the foundations of quantum mechanics.
In particular, the discussion on the minimal resources needed in order to obtain absolutely
secure
quantum communication is yet to be concluded.
Here we present an overview on our last experimental results concerning two novel quantum cryptographic schemes
 which do not require some of the most widely accepted conditions for realizing QKD.
The first is Goldenberg-Vaidman protocol \cite{v}, in which even if only orthogonal states
(that in general can be cloned without altering the state) are used, any eavesdropping attempt
is detectable.
The second is N09 protocol \cite{no} which, being based on the quantum counterfactual effect,
 does not even require any actual photon transmission in the quantum channel between the parties
 for the communication.
\end{abstract}

\section*{Introduction}
\addcontentsline{toc}{chapter}{Introduction}

Quantum Key Distribution (QKD) \cite{gis,scar} is beyond doubt the most promising application in the field of quantum technologies.
It is a method that allows two distant partners to share a common key to be used in order to perform a cryptographic communication whose
security does not depend on the technological level of an eventual spy, being based on the fundamental laws of nature and in particular on
the properties of quantum systems.

In the last decade QKD is moving from laboratories to become a
mature technology for commercialization \cite{l}; communications
over more than 100 km having been achieved both in fiber \cite{fib}
 and open air \cite{op}.

However, beyond its commercial interest QKD represents also a fruitful test bed of concepts and
ideas blossoming from quantum information theory and studies on foundations of quantum mechanics
\cite{gis,scar,mg,ekert,v,bostroem,lucamarini,shaari,maslennikov}.
In this paper, after a general introduction to the topic,  we present an overview on 
two novel QKD schemes, based on orthogonal states, which bring groundbreaking contribution in the debate on the actual resources needed
to ensure the security of quantum communication. Indeed, before the proposal of Goldenberg and Vaidman \cite{v} it was considered as established that non-orthogonal or entangled states were needed for QKD. This proposal, and later the one of T. G. Noh \cite{no}, demonstrated that non-orthogonality is not necessary, opening a new chapter for quantum communication, QKD with orthogonal states.
The paper is structured as follows: the first section after some brief historical information is addressed to the introduction of quantum cryptography and the description of the most studied protocols; the second section introduces the Goldenberg and Vaidman protocol based on the use of orthogonal states and describes its experimental implementation\cite{vnoi}; the third section concerns the N09 protocol as proposed by T. G. Noh, also dubbed Counterfactual QKD and its realization\cite{nonoi}; finally some final remarks and future perspectives are discussed.

\section{From classical to quantum cryptography}

In this section a brief introduction on classical and quantum cryptography is given. The aim is to follow the historical path that led to
current cryptographic systems and quantum cryptography.

\subsection{Classical Cryptography}
Cryptography concerns methods to encode messages so as to ensure privacy from anyone other than the authorized users.
Generally, the exchange of messages takes place between two characters named Alice (A) and Bob (B). A third character, Eve (E), attempts to intercept and decode the message.
The message to be sent from Alice to Bob is usually dubbed ``cleartext''.
To do this transmission safely Alice uses an algorithm (crypto-system or ``cipher'') able to combine the cleartext with some additional information (``key'') which is shared in an exclusive way between  A and B.
The text thus obtained is called ``cryptogram''.

Crypto-analysis, which is the study of how to decode a ciphered text without being in possession of the key, evolves in parallel to the development of cryptography.

Cryptography protocols can be divided into two main classes: symmetrical and asymmetrical. Protocols for which
the key used to encrypt the message is the same that is used to decode it, belong to symmetrical cryptography.
In asymmetric cryptographic systems, the key used to encrypt the message is different from that used to decode it.

\subsection{Symmetrical Key Distribution Protocols}
The cryptographic systems with symmetrical key are the most intuitive and oldest methods used to encode messages.



For example, one of the most ancient crypto-systems is the \textbf{Spartan scytale} 
The scytale is composed of a wood stick around which is coiled a tape. The result is a cylindrical surface to write.
Unrolling the tape and reading the message leads to a sequence of letters apparently without meaning. This sequence recovers
its original significance when the tape is rewound on a stick with the same diameter to the original one.
In this case the key of the protocol is the wood stick. The two parties must possess a wood stick of same diameter for exchange messages.

Another important cryptographic system has its origin back in Julius Caesar times. The \textbf{Caesar cipher} requires
 each letter that composes the message to be replaced with one that is three positions ahead in the alphabetic order as shown in Table \ref{Caesar}.
\begin{table}[htb]
\centering
\scriptsize{
\begin{tabular}{|c|c c c c c c c c c c c c c c c c c c c c c |}
 \hline
 Normal alphabet \hspace*{-0.25cm}&\hspace*{-0.25cm} A \hspace*{-0.25cm}&\hspace*{-0.25cm} B \hspace*{-0.25cm}&\hspace*{-0.25cm} C \hspace*{-0.25cm}&\hspace*{-0.25cm} D \hspace*{-0.25cm}&\hspace*{-0.25cm} E \hspace*{-0.25cm}&\hspace*{-0.25cm} F \hspace*{-0.25cm}&\hspace*{-0.25cm} G \hspace*{-0.25cm}&\hspace*{-0.25cm} H \hspace*{-0.25cm}&\hspace*{-0.25cm} I \hspace*{-0.25cm}&\hspace*{-0.25cm} L \hspace*{-0.25cm}&\hspace*{-0.25cm} M \hspace*{-0.25cm}&\hspace*{-0.25cm} N \hspace*{-0.25cm}&\hspace*{-0.25cm} O \hspace*{-0.25cm}&\hspace*{-0.25cm} P \hspace*{-0.25cm}&\hspace*{-0.25cm} Q \hspace*{-0.25cm}&\hspace*{-0.25cm} R \hspace*{-0.25cm}&\hspace*{-0.25cm} S \hspace*{-0.25cm}&\hspace*{-0.25cm} T \hspace*{-0.25cm}&\hspace*{-0.25cm} U \hspace*{-0.25cm}&\hspace*{-0.25cm} V \hspace*{-0.25cm}&\hspace*{-0.25cm} Z \\
 \hline
 Chiper alphabet \hspace*{-0.25cm}&\hspace*{-0.25cm} D \hspace*{-0.25cm}&\hspace*{-0.25cm} E \hspace*{-0.25cm}&\hspace*{-0.25cm} F \hspace*{-0.25cm}&\hspace*{-0.25cm} G \hspace*{-0.25cm}&\hspace*{-0.25cm} H \hspace*{-0.25cm}&\hspace*{-0.25cm} I \hspace*{-0.25cm}&\hspace*{-0.25cm} L \hspace*{-0.25cm}&\hspace*{-0.25cm} M \hspace*{-0.25cm}&\hspace*{-0.25cm} N \hspace*{-0.25cm}&\hspace*{-0.25cm} O \hspace*{-0.25cm}&\hspace*{-0.25cm} P \hspace*{-0.25cm}&\hspace*{-0.25cm} Q \hspace*{-0.25cm}&\hspace*{-0.25cm} R \hspace*{-0.25cm}&\hspace*{-0.25cm} S \hspace*{-0.25cm}&\hspace*{-0.25cm} T \hspace*{-0.25cm}&\hspace*{-0.25cm} U \hspace*{-0.25cm}&\hspace*{-0.25cm} V \hspace*{-0.25cm}&\hspace*{-0.25cm} Z \hspace*{-0.25cm}&\hspace*{-0.25cm} A \hspace*{-0.25cm}&\hspace*{-0.25cm} B \hspace*{-0.25cm}&\hspace*{-0.25cm} C\\
 \hline
\end{tabular}
}
\caption{Caesar chiper}
\label{Caesar}
 \end{table}

In this case the key of system is $3$, corresponding to the number of positions shifted. Obviously, with the same system is possible to use different keys.
A trivial example that illustrates the protocol is shown in Table \ref{Caesar2}.

\begin{table}[htb]
\centering
\scriptsize{
\begin{tabular}{|c| c c c c c c c c c c c|}
 \hline
 cleartext  \hspace*{-0.25cm}&\hspace*{-0.25cm} N \hspace*{-0.25cm}&\hspace*{-0.25cm} O \hspace*{-0.25cm}&\hspace*{-0.25cm} C \hspace*{-0.25cm}&\hspace*{-0.25cm} T \hspace*{-0.25cm}&\hspace*{-0.25cm} E \hspace*{-0.25cm}&\hspace*{-0.25cm} A \hspace*{-0.25cm}&\hspace*{-0.25cm} D \hspace*{-0.25cm}&\hspace*{-0.25cm} O \hspace*{-0.25cm}&\hspace*{-0.25cm} R \hspace*{-0.25cm}&\hspace*{-0.25cm} T \hspace*{-0.25cm}&\hspace*{-0.25cm} I\\
 \hline
 ciphertext \hspace*{-0.25cm}&\hspace*{-0.25cm} Q \hspace*{-0.25cm}&\hspace*{-0.25cm} R \hspace*{-0.25cm}&\hspace*{-0.25cm} F \hspace*{-0.25cm}&\hspace*{-0.25cm} Z \hspace*{-0.25cm}&\hspace*{-0.25cm} H \hspace*{-0.25cm}&\hspace*{-0.25cm} D \hspace*{-0.25cm}&\hspace*{-0.25cm} G \hspace*{-0.25cm}&\hspace*{-0.25cm} R \hspace*{-0.25cm}&\hspace*{-0.25cm} U \hspace*{-0.25cm}&\hspace*{-0.25cm} Z \hspace*{-0.25cm}&\hspace*{-0.25cm} N\\
 \hline
\end{tabular}
}
\caption{Example of application of Caesar cipher}
\label{Caesar2}
 \end{table}

Obviously, a cipher like this is very simple to decrypt as the key has only 20 possible values.
Caesar cipher and, in general, all methods for which a letter is encoded with the same symbol for the entire length of the message
are said to be based on  \textbf{monoalphabetic substitution}.
This type of cryptographic systems can be decrypted by applying a technique known as frequency analysis.
The first known recorded explanation of frequency analysis (indeed, of any kind of cryptanalysis) was given in the 9th century by Al-Kindi, an Arab polymath,
in ``A Manuscript on Deciphering Cryptographic Messages''.
It has been suggested that close textual study of the Qur'an first brought to light that Arabic has a characteristic letter frequency.
Frequency analysis is based on the fact that, in any given stretch of written language, certain letters and combinations of letters occur with varying frequencies.
Moreover, there is a characteristic distribution of letters that is roughly the same for almost all samples of that language.
For instance, given a section of Italian language, E, A, I and O are the most common, while Z, Q and V are rare.
So by analyzing a message, coded with the monoalphabetic substitution, one can obtain the frequency spectrum of each symbol. Comparing it
with the one relative to the letters of a given language one can derive the key to decrypt the message.

The natural evolution of this cryptographic system is the use of more than one alphabet for encoding the message.
The key will be used to indicate which alphabet to use for each letter.
This type of cryptographic system is known as \textbf{polyalphabetic substitution}.
The first complete description of a cryptographic system based on polyalphabetic substitution dates back to Leon Battista Alberti (circa 1467),
who used a metal cipher disc to switch between cipher alphabets. Alberti's system only switched alphabets after several words, and switches
were indicated by writing the letter of the corresponding alphabet in the ciphertext.

Years later, in 1508, Johannes Trithemius invented the \textbf{Tabula Recta}.
The tabula recta is a square table of alphabets, each row of which is made by shifting the previous one to the left.
In his system data are encrypted by switching each letter of the message with the letter directly below, using the first shifted alphabet.
The next letter is switched by using the second shifted alphabet, and this continues until the end of the message.

Around 1586 Blaise de Vigen\`{e}re developed a more secure system always based on Tabula Recta.
This protocol is called \textbf{Vigen\`{e}re cipher}.
The difference from the previous one is that the sequence of alphabets used to encrypt the message was not
consecutive, but dictated by a word or phrase: the key.
Each letter of the key is matched with a letter of the message, the key is repeated along the length of the message.
In Table \ref{vigenere} is shown a simple example on the operation of the Vigen\`{e}re cipher, where the keyword is: ``GREEN''.

\begin{table}[htb]
\centering
 \scriptsize{
 \begin{tabular}{|c|c c c c c c c c c c c c c c c c c c c c c c c c|}
  \hline
  Cleartext &\hspace*{-0.25cm} K \hspace*{-0.25cm}&\hspace*{-0.25cm} I \hspace*{-0.25cm}&\hspace*{-0.25cm} L \hspace*{-0.25cm}&\hspace*{-0.25cm} L \hspace*{-0.25cm}&\hspace*{-0.25cm} K \hspace*{-0.25cm}&\hspace*{-0.25cm} I \hspace*{-0.25cm}&\hspace*{-0.25cm} N \hspace*{-0.25cm}&\hspace*{-0.25cm} G \hspace*{-0.25cm}&\hspace*{-0.25cm} T \hspace*{-0.25cm}&\hspace*{-0.25cm} O \hspace*{-0.25cm}&\hspace*{-0.25cm} M \hspace*{-0.25cm}&\hspace*{-0.25cm} O \hspace*{-0.25cm}&\hspace*{-0.25cm} R \hspace*{-0.25cm}&\hspace*{-0.25cm} R \hspace*{-0.25cm}&\hspace*{-0.25cm} O \hspace*{-0.25cm}&\hspace*{-0.25cm} W \hspace*{-0.25cm}&\hspace*{-0.25cm} M \hspace*{-0.25cm}&\hspace*{-0.25cm} I \hspace*{-0.25cm}&\hspace*{-0.25cm} D \hspace*{-0.25cm}&\hspace*{-0.25cm} N \hspace*{-0.25cm}&\hspace*{-0.25cm} I \hspace*{-0.25cm}&\hspace*{-0.25cm} G \hspace*{-0.25cm}&\hspace*{-0.25cm} H \hspace*{-0.25cm}&\hspace*{-0.25cm} T\\
  \hline
  Key       &\hspace*{-0.25cm} G \hspace*{-0.25cm}&\hspace*{-0.25cm} R \hspace*{-0.25cm}&\hspace*{-0.25cm} E \hspace*{-0.25cm}&\hspace*{-0.25cm} E \hspace*{-0.25cm}&\hspace*{-0.25cm} N \hspace*{-0.25cm}&\hspace*{-0.25cm} G \hspace*{-0.25cm}&\hspace*{-0.25cm} R \hspace*{-0.25cm}&\hspace*{-0.25cm} E \hspace*{-0.25cm}&\hspace*{-0.25cm} E \hspace*{-0.25cm}&\hspace*{-0.25cm} N \hspace*{-0.25cm}&\hspace*{-0.25cm} G \hspace*{-0.25cm}&\hspace*{-0.25cm} R \hspace*{-0.25cm}&\hspace*{-0.25cm} E \hspace*{-0.25cm}&\hspace*{-0.25cm} E \hspace*{-0.25cm}&\hspace*{-0.25cm} N \hspace*{-0.25cm}&\hspace*{-0.25cm} G \hspace*{-0.25cm}&\hspace*{-0.25cm} R \hspace*{-0.25cm}&\hspace*{-0.25cm} E \hspace*{-0.25cm}&\hspace*{-0.25cm} E \hspace*{-0.25cm}&\hspace*{-0.25cm} N \hspace*{-0.25cm}&\hspace*{-0.25cm} G \hspace*{-0.25cm}&\hspace*{-0.25cm} R \hspace*{-0.25cm}&\hspace*{-0.25cm} E \hspace*{-0.25cm}&\hspace*{-0.25cm} E\\
  \hline
  Cryptogram&\hspace*{-0.25cm} Q \hspace*{-0.25cm}&\hspace*{-0.25cm} Z \hspace*{-0.25cm}&\hspace*{-0.25cm} P \hspace*{-0.25cm}&\hspace*{-0.25cm} P \hspace*{-0.25cm}&\hspace*{-0.25cm} X \hspace*{-0.25cm}&\hspace*{-0.25cm} O \hspace*{-0.25cm}&\hspace*{-0.25cm} E \hspace*{-0.25cm}&\hspace*{-0.25cm} K \hspace*{-0.25cm}&\hspace*{-0.25cm} X \hspace*{-0.25cm}&\hspace*{-0.25cm} B \hspace*{-0.25cm}&\hspace*{-0.25cm} S \hspace*{-0.25cm}&\hspace*{-0.25cm} F \hspace*{-0.25cm}&\hspace*{-0.25cm} V \hspace*{-0.25cm}&\hspace*{-0.25cm} V \hspace*{-0.25cm}&\hspace*{-0.25cm} B \hspace*{-0.25cm}&\hspace*{-0.25cm} C \hspace*{-0.25cm}&\hspace*{-0.25cm} D \hspace*{-0.25cm}&\hspace*{-0.25cm} M \hspace*{-0.25cm}&\hspace*{-0.25cm} H \hspace*{-0.25cm}&\hspace*{-0.25cm} A \hspace*{-0.25cm}&\hspace*{-0.25cm} O \hspace*{-0.25cm}&\hspace*{-0.25cm} X \hspace*{-0.25cm}&\hspace*{-0.25cm} L \hspace*{-0.25cm}&\hspace*{-0.25cm} X\\
  \hline
 \end{tabular}
}
 \caption{Vigen\`{e}re cipher}
 \label{vigenere}
\end{table}

For many years this cryptographic system was considered impossible to violate. In 1863 the first publication on an attack at the Vigen\`{e}re cipher was realized.
This method, known as Kasiski examination, allows to determine the length of the key; then one can apply the frequency analysis on each alphabet.
The Kasiski examination consists in the research in the encrypted text of equal strings. If these are encoded with the same part of the key then
the distance between the parts must be an integer multiple of the key length.
Moreover, the key was usually a word or a phrase whereby is not necessary to earn each letter with the frequency analysis. For example if one obtained from
the analysis only a few letters of the key as ``G*EE*'' is not difficult to understand that the keyword is ``GREEN''.

In 1917 Gilbert Vernam, an engineer at Bell Labs, modified the Vigen\`{e}re cipher to make it more secure.
The Vernam cipher is also called \textbf{One-Time Pad}.
This cryptographic system is based on three conditions:
\begin{itemize}
 \item the key must have the same length of the message
 \item it must be used only one time
 \item it must be randomly generated
\end{itemize}
The condition on the key length makes the single message completely immune from the frequency analysis.
Notwithstanding that the same type of alphabet is used for encoding more than one letter it is impossible to determine where this alphabet is used.
The second condition addresses the issue that if the key is used more than one time, in principle, Eve can collect all messages and then apply the frequencies analysis on the same position of all texts.
Finally, if the key is randomly generated no correlation between successive alphabets can be found, at variance with  the case of Vigen\`{e}re cipher.
After the end of World War II (1949) Claude Shannon demonstrated that One-time pad is unbreakable \cite{shannon}.
With the development of computer science all communications have been converted to digital encoding, so both the key and the message are now composed by digital bits.
As a consequence, all cryptographic systems are now based on binary operations. For example for encoding messages the logic XOR operation is used between
the key and the message bits.
The three conditions necessary for the cipher to work present different practical difficulties.
In fact, the problem of the frequent transmission of the key bits (which have to be strictly random) through a secure channel leads to the main technological issue related to modern cryptography (key distribution).
However this problem can in principle be solved with the use of quantum cryptography as we shall address in section \ref{sec:QKD}.

\subsection{Asymmetrical Key Distribution}
The second class of cryptographic protocols is composed by asymmetrical cryptosystems, more commonly called public key cryptosystems.
The fundamental characteristic of these protocols is that the key used for encoding the message (called ``public key'')
is different from the one used to decode it (called ``private key'').
The public key is exchanged between Alice and Bob through a public channel and it could be, in principle, intercepted by Eve.
On the contrary, the private key remains always in possession of Alice and Eve cannot get hold of it in any way.
Another fundamental characteristic of this kind of protocols is the possibility of establishing a shared secret-key
over an authenticated (but not private) communications channel without using a prior shared secret.
To give an idea of how the asymmetric protocols work, suppose that Bob wants to send a message to Alice.
Imagine that Alice sends  Bob a box  and an open padlock (public key).
Bob puts the message in the box, closes it with the padlock and sends the box to Alice. Alice can open the
box with the key of the padlock that has always remained only in her possession (private key).
In principle, a copy of the padlock can be in possession of Eve but nevertheless she can not open the box in any way.
In general this type of protocol is implemented by using a class of mathematical functions called one-way functions.
These functions have the characteristic of being easily computed with any input but they are hard to invert.
The terms ``easy'' and ``hard'' are in this context to be intended in terms of computational time requested to solve the given problem.
In particular, a function can be considered  ``easy'' to compute if the evaluation time is proportional to a power of the length of the input string (polynomial class).
On the other hand, a function is said ``hard'' to compute if the time is instead an exponential function of the length of the input string.

One of the unsolved problems in computer science is to prove the existence of these functions:
all the functions that are  generally included in this class are actually the ones for which an algorithm to solve them in polynomial time has not been found yet, but it is not proved that such an algorithm does not exist.
In terms of functions Alice builds the public key $f(x)$ and the private key $f^{-1}(x)$. She sends $f(x)$ to Bob.
Bob computes the function with the message $M$ and he obtains $f(M)$. The fundamental concept is that obtaining $f^{-1}(x)$ only with $f(x)$ as a starting point is a ``hard'' task.
Finally Bob sends $f(M)$ to Alice and than she computes $f^{-1}(f(M))=M$ retrieving the message.

The most important public-key protocol is RSA, originally
invented in 1978 by Rivest, Shamir and Adleman.
The security of the protocol is based on the apparent difficulty of the integer factorization problem.
The integer factorization is a much discussed problem. To date, the best way to solve it is the use of a class $O(exp((\frac{64}{9}L)^{\frac{1}{3}}(log L)^{\frac{2}{3}}))$ algorithm,
where $L$ is the length of the key. So this problem is ``hard'' to resolve as it need exponential computational time.
Even if RSA is the most used cryptosystem for commercial use,
however, there is no proof that there cannot be more efficient algorithms.
Based on current algorithms and computational power of computers, to ensure the security of the protocol, the key must be at least 2048 bits.
In 1994, Peter Shor showed that a quantum computer would be able to solve factorization problem in polynomial time, thus giving the possibility to break RSA protocol \cite{shor}.

\subsection{Quantum Key Distribution}
\label{sec:QKD}

Quantum states present some particular properties which can be exploited to perform a secure exchange of information between two parties. In particular to obtain a secure shared key to be used in a one-time-pad protocol. For this reason Quantum Cryptography was introduced or, more properly, Quantum Key Distribution (QKD).

The first idea of Quantum Cryptography was proposed by Stephen Wiesner, at Columbia University in New York in the early 1970s.
He introduced the concept of quantum conjugate coding. Initially his work was refused by IEEE Information Theory and it was
published only in 1983 by SIGACT News (Special Interest Group on Algorithms and Computation Theory) \cite{conjugatecoding}.
About one year later Charles H. Bennet and Gilles Brassard
announced a protocol of quantum cryptography based on non-orthogonal states \cite{BB84}.

The most of the Quantum Cryptography protocols are based on the No-Cloning theorem.

This theorem asserts that it is impossible to copy exactly (cloning) an a priori unknown quantum state.
However, cloning is possible if the state belongs to a set of orthogonal states known.
This is important considering a scenario in which A and B decide to exchange information coded in the states of quantum systems. The most simple strategy that E can use is to intercept the state sent by A before it reaches B, to measure it, to copy it and then send the copy to Bob (intercept-resend attack).
Let us consider two quantum states $\mid \psi \rangle$ e $\mid \phi \rangle$ to be cloned and an auxiliary state $\mid s \rangle$ which is intended to become the clone (target state).

\begin{equation}
 \mid \psi \rangle \mid s \rangle \rightarrow \mid \psi \rangle \mid \psi \rangle
\end{equation}
If it is possible to perform the cloning then a unitary operator $U$ must exist  such that:
$$U(\mid \psi \rangle \mid s \rangle)=\mid \psi \rangle \mid \psi \rangle$$
$$U(\mid \phi \rangle \mid s \rangle)=\mid \phi \rangle \mid \phi \rangle$$
and by taking the scalar product of the two equation one would obtain:
\begin{equation}
 \langle \psi \mid \phi \rangle = \langle \psi \mid \phi \rangle \langle \psi \mid \phi \rangle
\end{equation}
For the former equation to be true one should have
$\langle \psi \mid \phi \rangle$ equal either to $1$ or $0$, meaning that  $\mid \psi \rangle$ and $\mid \phi \rangle$ are equal or orthogonal respectively.
In any other case $U$ does not exist so it is not possible to clone an {\em a priori} unknown quantum state. This means that if E tries to perform an intercept-resend attack, the states sent by E to B will be uncorrelated to the original states generated by A, leading to errors in the transmission that can highlight the presence of the eavesdropper.

The first and most important QKD protocol was proposed in 1984 by Charles Bennet and Gilles Brassard during a conference in India \cite{BB84}.
This protocol involves the use of polarized photons in two non-orthonormal bases: one rectilinear (
$\mid \leftrightarrow \rangle$ e $ \mid \; \updownarrow\; \rangle$) and one diagonal ($\mid \nearrow \rangle \mid \searrow \rangle$).
Suppose that for each base the first state correspond to the bit $0$ and the second to bit $1$.
The protocol can be implemented as follows:

\begin{itemize}
 \item key transmission: Alice prepares a random sequence of bits  (Table \ref{BeBr84} a) and a random sequence of bases (choosing between the rectilinear and the diagonal one) (\ref{BeBr84} b) to encode them. Bob prepares a random sequence of bases to perform the measurement \ref{BeBr84} d). Alice then sends  a succession of linearly polarized photons according to the chosen bits and bases \ref{BeBr84} c). Bob projects each photon according to the chosen bases and stores the results of the measurements;
\item public discussion: Alice e Bob announce on the public channel the sequence of chosen bases. Bob keeps the results related to the measurements with compatible bases and discard the others. Alice also keeps the bits corresponding the states for which the same basis has been used. The sequences of bits obtained in this way by A and B are now identical (assuming no noise and no eavesdropping). This is to be used as cryptographic key;
\item error detection: To check for the eventual presence of E in the communication channel, A and B sacrifice random portions of the shared string and publicly compare their positions and values (or the parities of corresponding subsets) before discarding them. The remaining bits are called sifted key.
\end{itemize}
%
%
%
\begin{table}[htb]
\centering
\scriptsize{
\begin{tabular}{|c c c c c c c c c c c c c c c c|}
\hline
 a)...& 0 & 1 & 1  & 0 & 1  & 1 & 0 & 0  & 1 & 0 & 1 & 1  & 0 & 0  & 1  \\
 b)...& D & R & D  & R & R  & R & R & R  & D & D & R & D  & D & D  & R  \\
 c)...& $\nearrow$ & $\updownarrow$ & $\searrow$ & $\leftrightarrow$ & $\updownarrow$ & $\updownarrow$ & $\leftrightarrow$ & $\leftrightarrow$ & $\searrow$ & $\nearrow$ & $\updownarrow$ & $\searrow$ & $\nearrow$ & $\nearrow$ & $\updownarrow$ \\
 d)...& R & D & D  & R & R  & D & D & R  & D & R & D & D  & D & D  & R  \\
 e)...& 1 &   & 1  &   & 1  & 0 & 0 & 0  &   & 1 & 1 & 1  &   & 0  & 1  \\
 f)...& R &   & D  &   & R  & D & D & R  &   & R & D & D  &   & D  & R  \\
 g)...&   &   & ok &   & ok &   &   & ok &   &   &   & ok &   & ok & ok \\
 h)...&   &   & 1  &   & 1  &   &   & 0  &   &   &   & 1  &   & 0  & 1  \\
 i)...&   &   &    &   & 1  &   &   &    &   &   &   &    &   & 0  &    \\
 l)...&   &   &    &   & ok &   &   &    &   &   &   &    &   & ok &    \\
 m)...&   &   & 1  &   &    &   &   & 0  &   &   &   & 1  &   &    & 1  \\
\hline
\end{tabular}
}
\caption{Scheme of BB84 protocol}
\label{BeBr84}
\end{table}

In the following years many experimental implementations of BB84 have been performed and several alternative protocols have been proposed, almost all of them based on the use of non-orthogonal states. Since this work is not intended to
give an exhaustive review on that kind protocols, but rather introduce experiments which do not require such states, we refer on the subject to other important works \cite{gis,scar,mg,ekert,bostroem,lucamarini,4,5,6,7,8,9,10,11,12,13,14,15,16,17,18}.
Let now us mention just another possible attack that can be performed by Eve.
Eve's possible attacks discussed above are based on the attempt to get the maximum information out of the qubits exchanged between Alice and Bob.
Eve can also adopt a completely different strategy: she can send herself signals in the quantum channel toward Alice's or Bob's secure zones.
This kind of strategies are called Trojan horse attacks. Eve can send light pulses in the fibers entering in Alice's and Bob's systems and analyze the back-reflected light.
In principle, with this analysis, is possible to determine which detector just fired, 
which laser just flashed and which polarization is set.
However, the back-reflected light is extremely low so this type of attack is difficult to perform. Eve, to avoid being revealed, can send light pulses with a wavelength
completely different to that used for the protocol and in which A and B detectors are inefficient. This strategy can be prevented by the use of filter with a transmission
spectrum corresponding to the sensitivity of the detectors.
The mere fact that this type of attacks exists makes it clear that the security of QC cannot be guaranteed only by the use of quantum mechanic properties but also requires technological countermeasures.


\section{Goldenberg-Vaidman protocol}
In the following we introduce our first implementation of an experiment concerning
QKD featuring orthogonal states.

\subsection{GV protocol: the scheme}
In the proposal \cite{v}, the orthogonal states sent by Alice are the
superpositions of two localized wave-packets, which do not travel simultaneously to Bob, being
separated by a fixed delay.
There is a direct correspondence between the state
prepared by Alice and the bit received by Bob, for instance
\begin{eqnarray}
\label{eq:states}
0  &\rightarrow \ket{\Psi_0}&=\frac{1}{\sqrt{2}}(\ket{a}+\ket{b})\nonumber \\
1 &\rightarrow \ket{\Psi_1}&=\frac{1}{\sqrt{2}}(\ket{a}-\ket{b}),\nonumber
\end{eqnarray}
where $\ket{a}$ and $\ket{b}$ are two localized wave--packets and
the states $\ket{\Psi_0}$ and $\ket{\Psi_1}$ are orthogonal. The
states $\ket{\Psi_0}$ and $\ket{\Psi_1}$ are emitted randomly in
time, and the presence of an eventual eavesdropper can be detected
by legitimate users exploiting the information on the detection
times \cite{v}.

The protocol works as follows: Alice sends Bob either the state $\ket{\Psi_0}$ or $\ket{\Psi_1}$.

The insertion on the quantum channel of the wave-packet $\ket{b}$ is delayed for some amount of time
$\tau$ with respect to the insertion of wave-packet $\ket{a}$. $\tau$ is chosen larger than the
traveling time $T$ of photons between Alice's and Bob's locations. Since $\ket{b}$  travels through
the quantum channel only after the wave-packet $\ket{a}$ has already reached Bob's site, both
wavw-packets are never simultaneously present in the quantum channels. Nonetheless, the requirement of $\tau$ greater than the traveling time $T$ is not strictly
necessary. 

\cite{v}.

Our proof-of-principle experiment\cite{vnoi}  exploits a
balanced Mach-Zehnder Interferometer (MZI) with two equal optical delays
$OD_{1}$ and $OD_{2}$. According to Fig. 1, sources of single photon
$S_{0}$ and $S_{1}$ at the two input ports of the beam splitter on
Alice side provide single photons propagating in the transmission
channel in the state $\ket{\Psi_0}$ or $\ket{\Psi_1}$ respectively.
The emission time of the single photon in one of the two states is absolutely
random, but it is registered by Alice.

while the $\ket{b}$ is stored in $OD_{1}$, wave-packet $\ket{a}$ travels from Alice's to Bob's
locations along the upper channel and enters in $OD_{2}$, where it is delayed until also $\ket{b}$
reaches Bob's site. Thus the two packets interfere as they simultaneously  reach the
second beam-splitter. A click of detector $D_{i}$ deterministically implies that the single
photon state was in the state $\ket{\Psi_{i}}$, that is, it was sent by source $S_i$.
Two security tests are performed by Alice and Bob to highlight the possible presence of an
eavesdropper. The first is a public comparison between the sending times ${t_{s}}$ and the
receiving times ${t_{r}}$ for each photon. If we suppose that the traveling time between the two
parties is $T$, only the events detected at time $t_{r}=t_{s}+\tau+T$ are considered
as part of the message,
 while all the
others,
highlighting the presence of Eve, are discarded. The second one is the comparison of corresponding segments of the
users' bit strings to estimate the quantum bit error rate (QBER). We underline that in the ideal
case discrepancies in the transmission/detection times or in the bit strings can only be induced by
Eve.

For the sake of completeness, let us mention that it was argued by Peres \cite{peres}
that this protocol introduced no novel features with respect to BB84.  Goldenberg and Vaidman
replied to this claim by stating that while in other protocols (like BB84) the security is obtained by virtue of non-orthogonality,
in GV protocol it is due to causality, since they proved that super-luminal signaling would be required for a successful eavesdropping \cite{v}.
Furthermore, while all cryptographic schemes require two steps for sending information
(sending the quantum object and then some classical information), in GV protocol
only the first step is needed for communication, the second step being used only for
ensuring security against eavesdropping.

\subsection{GV protocol: the experimental setup}

\begin{figure}
\begin{center}
\includegraphics[scale=0.25, angle=90]{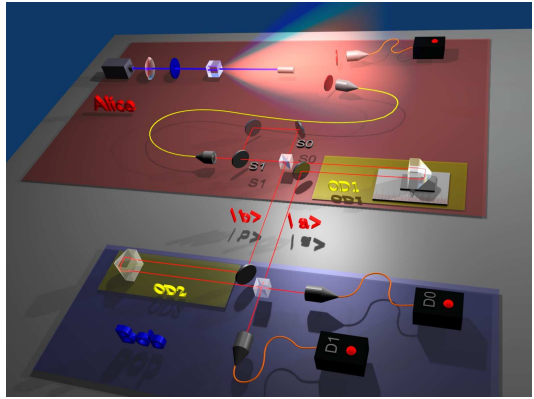}
\end{center}
\caption{ \label{fig:example2} Experimental GV protocol set-up. A single photon source (realized exploiting an heralded single photon source based on parametric down-conversion (PDC) obtained by pumping with a $406$ nm CW laser beam a type I BBO crystal) can be injected deterministically in either of the two input ports ($S_0$, $S_1$) of a balanced Mach-Zehnder Interferometer (MZI) , encoding (respectively) bit $0$ or $1$. The choice is random and performed by passive optics.
Alice's site is composed by $S_0$, $S_1$ and the first optical delay ($OD_{1}$).
Bob's location is composed by the second optical delay ($OD_{2}$, identical to $OD_{1}$)
and the two single photon detectors ($D_{0}, D_{1}$).}
\end{figure}

The setup of the experiment \cite{vnoi} representing the first realization of the GV
protocol is shown in Fig. 1. The single photon states are obtained exploiting an heralded single photon source based
on parametric down-conversion (PDC) \cite{mandel}. A CW $100$ mW Coherent Cube diode laser system
at $406$ nm is used to pump a BBO type I crystal. PDC photons pairs at degeneracy ($812$ nm) are
emitted in slightly non-collinear regime (three degrees with respect to the pump direction). The
heralding photons are selected by means of $1$ nm bandwidth interference filters, collected in a
multimode optical fiber and detected by Single Photon Avalanche Photo-diodes (SPAD) detectors. The
heralded single photon, the carrier of the information to be exchanged between Alice and Bob, is collected in a single mode optical fiber (a $10$ nm interference filter is placed on
the heralding arm only for background suppression). The CW laser operation ensures the generation
of photon pairs at random time, and the detection of one photon of the pair in the heralding arm
provides the temporal information on the emission of the single-photon, as requested by the original proposal.
To perform the proof-of-principle of the QKD scheme, Alice sends bit
$0$ or $1$ by addressing the encoding photon to the proper input port of the MZI 
($S_0$ or $S_1$ respectively). In our case, this can be  achieved just switching the optical fiber
from one input port to another. It is noteworthy to observe that in practical QKD systems this can be realized
exploiting a commercial fast optical switch controlled by a random number generator.
Bob detects the single photons at the output of the interferometer.
The balanced MZI 
contains both the optical delays and the transmission
channel between the users. In particular, after the input BS at the Alice side one arm of the
interferometer contains a delay line, while on the other arm the
delay line (both delays are based on trombone prisms) is located at Bob's side. The positions of the
trombones in the optical delays are adjusted via a closed loop piezo-movement system with
nano-metric resolution. Detection events after the output BS of the interferometer are revealed by
SPAD detectors operating in Geiger mode. The electronics highlighting the presence of coincident
detections is based on Time-to-Amplitude-Converter and Multi-Channel-Analyzer.
In our case the temporal condition for the security of the QKD scheme is satisfied because
the jitter of our detectors (i. e., the uncertainty in the measurement of the
transmission/detection times) is about 300 ps, while the length of the delay lines is approximately 60 cm
corresponding to a storage time of $\sim$2 ns. Furthermore, as the signal
corresponding to the detection events on the heralding channel (containing the information on  $t_s$) is properly delayed before exiting from A site, Eve can never access
the timing information before the transmission of each photon is concluded.

The stability of the interferometer has been tested by scanning the position of Alice's trombone
prism with Bob's one kept at a fixed position. Fig. 2 shows the interference fringes of heralded
counts. The visibilities (V) are well above $80 \%$, irrespective of which port of the input
beam splitter is used to inject the single photon in the interferometer. Even if, in recent years,
very high visibilities have been achieved in similar setups \cite{nc1,nc2},
the results we obtain are absolutely comparable with those of several important works \cite{c1,c2,c3,c4} 
and they are absolutely sufficient for a meaningful proof-of-principle of the GV protocol.
\begin{figure}
   \begin{center}
   \includegraphics[width=9cm]{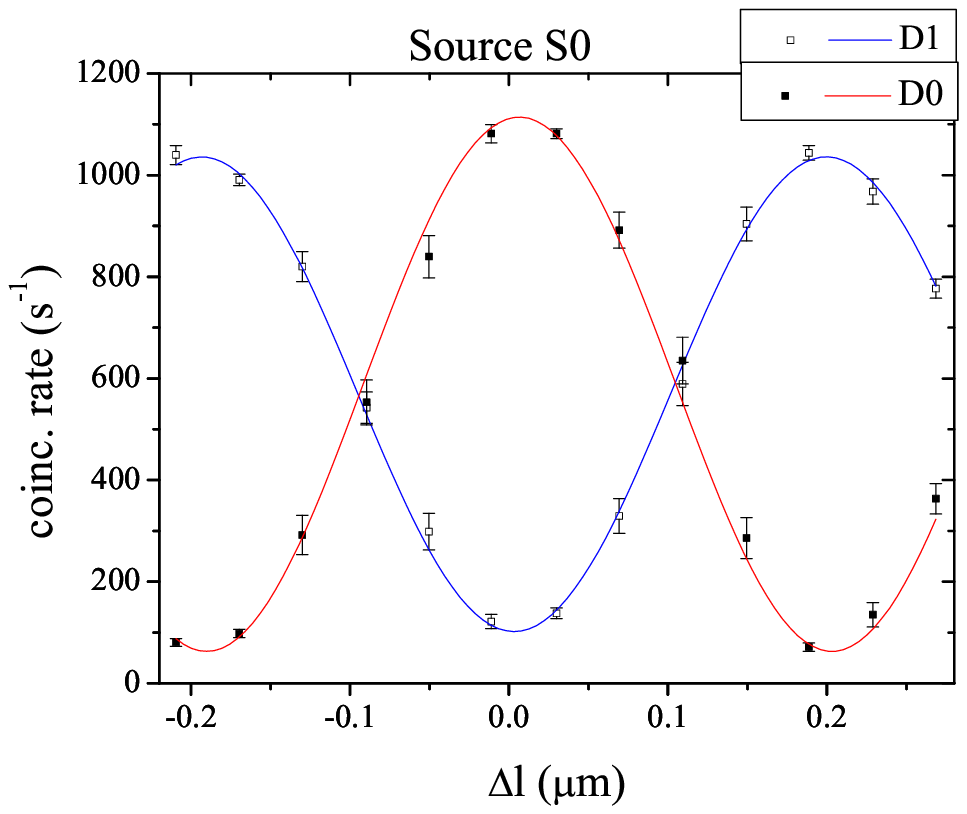}\\ \includegraphics[width=9cm]{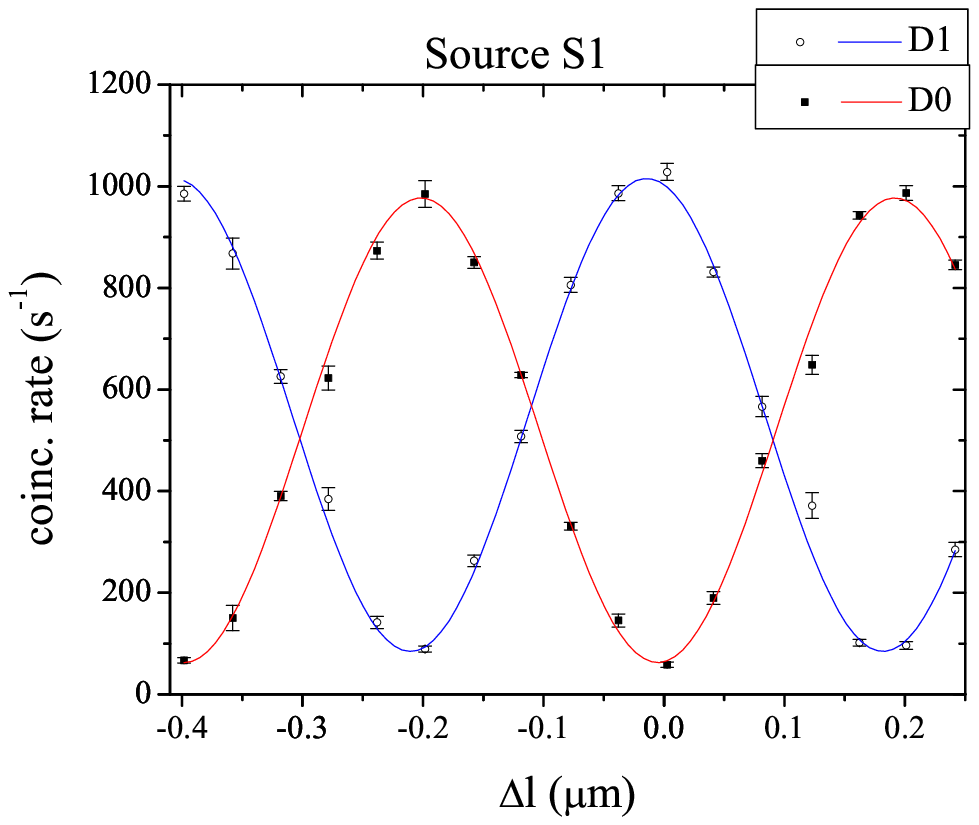}
   \end{center}
   \caption[example5]
   { \label{fig:example5}  (Color online) Number of detected events per second at detector
$D_0$ and $D_1$ 
as a function of the path
length difference $\Delta l$ 
between the two arms of the
interferometer for source $S_0$ (top picture) and $S_1$ (bottom). 
visibility of the observed fringes is reported in Tab $1$.}
   \end{figure}

\begin{figure}
   \centering
   \begin{center}
    \includegraphics[width=9cm]{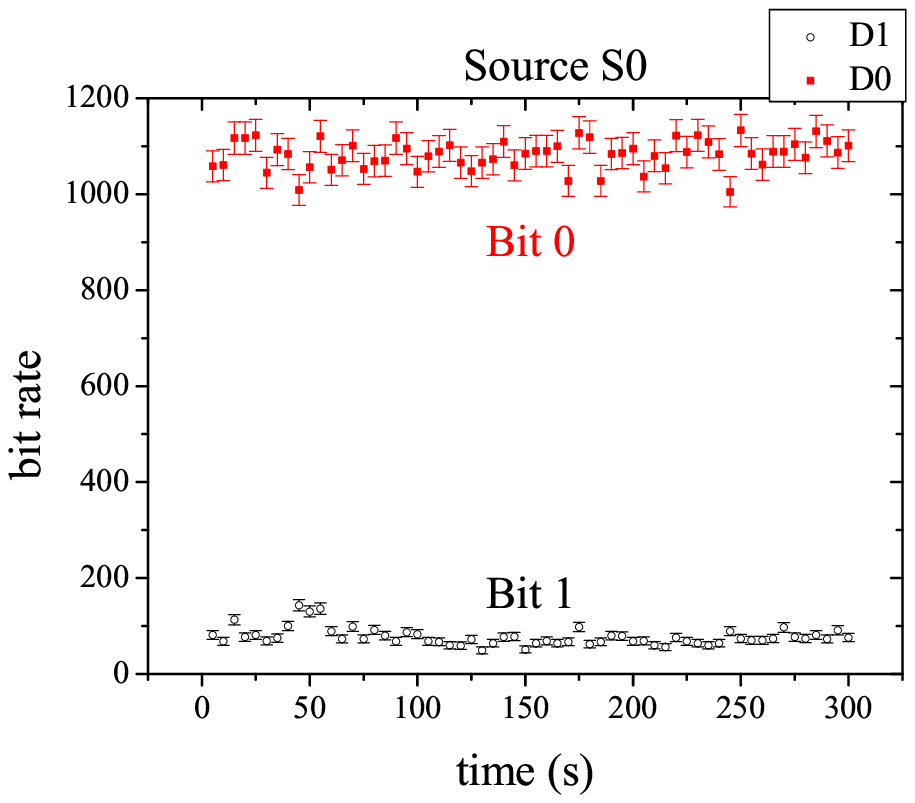}\\  \includegraphics[width=9cm]{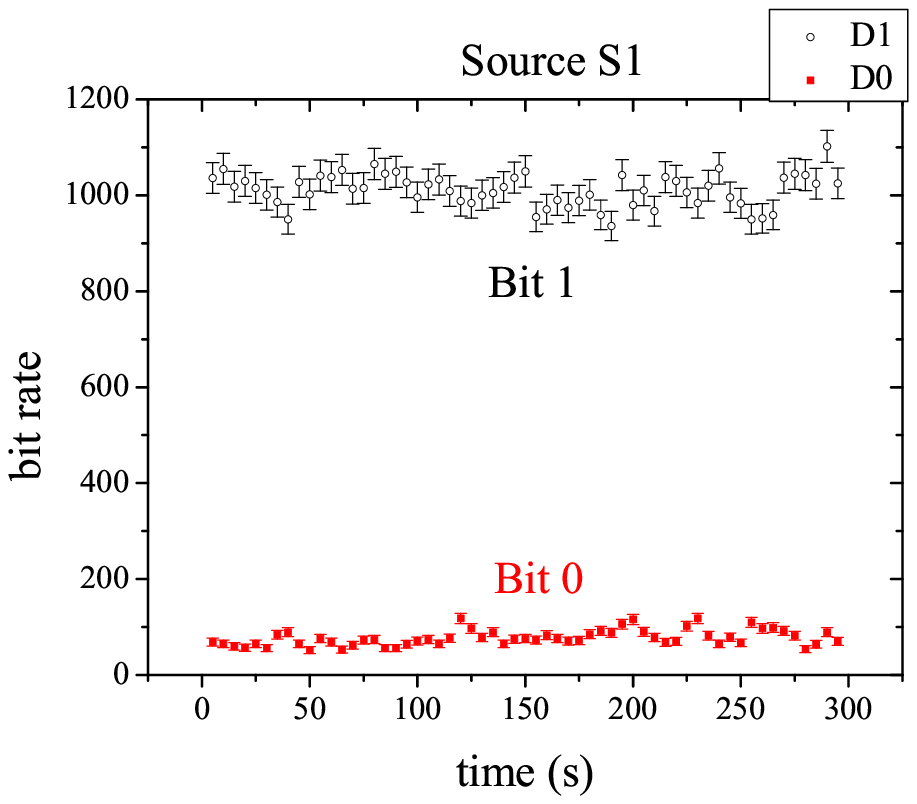}
   \end{center}
   \caption{\label{fig:example6}  Detection events at both detectors $D_1$ and $D_0$. Top: source $S_0$ is active,
   corresponding to the transmission of a string of bit $0$. Bottom: source $S_1$ is active, corresponding
 to the transmission of a string of bit $1$. The evaluated Quantum Bit Error Rate ($QBER$) in the two
   cases are $QBER_{S1}=0.071 \pm 0.014$ and $QBER_{S0} = 0.070 \pm 0.016$ on a series of $60$ measurements
   5 seconds long, showing a remarkable phase stability of the interferometer.}
   \end{figure}

\subsection{Results and discussion}

The quality of the transmission is quantified by the Quantum Bit Error Rate
($QBER=\frac{P_{Wrong}}{P_{Right}+P_{Wrong}}$, where $P_{Right}$ ($P_{Wrong}$) is the probability for Bob to receive a bit value which is equal to (different from) the one sent by Alice), measured to be $7\%$. This result has been proven to be stable for hundreds of seconds as  shown
 in Fig. 3. The main results of our transmission are summarized in Table I. 

\begin{table}[htbp]
\centering
   \begin{tabular}{|c|c|c|c|}
   \hline
    & $V_{D0}$ & $V_{D1}$ & $QBER$\\
   \hline
   $S0$ & $(89 \pm 1)\%$ & $(82 \pm 1)\%$ & $(7.0 \pm 1.6)\%$\\
   \hline
   $S1$ & $(88 \pm 1)\%$ & $(85 \pm 1)\%$ & $(7.1 \pm 1.4)\%$\\
   \hline
   \end{tabular}
\caption{Main results obtained in our implementation of the QKD protocol proposed in Ref. \cite{v}.
$V_{D0}$, $V_{D1}$ are the visibilities of the interference fringes observed at the two outputs of the interferometer by scanning the path length difference, $QBER$ is the estimated quantum bit error rate
for the transmission
. }
   \end{table}

Finally, some observations regarding the security of this QKD system. Despite the fact that an unconditional security proof of the GV protocol is still not available, we note that an efficient eavesdropping strategy against its ideal realization has not been found yet. On the contrary, it can be shown that, if a multi-photon component is present in the signal,
an eavesdropper (Eve) can gain information on the key by performing a beam-splitter attack. For example, Eve can insert a
beam-splitter, in both paths 
in such a way that the transmitted photons continue traveling toward Bob while measuring the outputs in the reflected modes 
with a duplicated Bob's detection apparatus.
In order to successfully do this Eve should be present since the initial tuning of the interferometer. 
The security issue in QKD protocols based on single photons due to the presence of multi-photon components is a very deeply investigated subject \cite{gis,scar}, which demands ultimately the use of efficient
single photon sources.
In particular, our heralded single photon source, presenting a $g^{(2)}(0)=0.06 \pm 0.01$, is a good approximation of an ideal single photon source, thus the information obtained by an eventual eavesdropper exploiting the presence of multi-photon components is negligible. In fact, if we attribute the measured QBER value, due to experimental imperfections, to an attack performed by an eavesdropper, the amount of information on the key obtained by this attack will be much greater than the one obtained from a beam-splitting attack on our ``almost ideal single photons''.

\section{Counterfactual Quantum Cryptography}

In the following we describe the first implementation of
a counterfactual QKD experiment (Noh09 Protocol). After the introduction of the mechanism of the counterfactual
measurement via the description of famous Elitzur-Vaidman bomb tester we briefly describe the theoretical
proposal, we show the details of the implementation, we present the experimental data and discuss the results.

\subsection{Elitzur-Vaidman bomb tester}
The Elitzur-Vaidman bomb tester \cite{BOMB1} is a {\em gedanken experiment} that shows the peculiar quantum counterfactual effect.
Consider a collection of bombs, some of which are duds while the remaining ones are usable. The bombs have a particular property:
each bomb has a perfect photon-triggered sensor which whenever a photon is absorbed causes the bomb to detonate. We assume that
the efficiency of this sensor is 100$\%$ so that all photons that pass through the sensor will be absorbed.
The dud bombs has a malfunctioning sensor that will not interact with any photon. The problem is how to separate
dud bombs from usable ones without detonating all bombs.
For solving this problem the authors proposed a mode of observation known as counterfactual measurement or interaction-free measurement.
Basically the bomb to analyze is positioned on one arm of a Mach-Zehnder interferometer. 
If there is not any bomb,  the photons, due to single photon interference, are detected always from the same detector.
Assume now that this detector is dubbed detector A and that a bomb is introduced. If the bomb is dud nothing changes because it does not interact with photons, so they are still detected by detector A.
If the bomb is usable, the interference is broken. In fact, if a photon passes on the lower branch it will be absorbed by the bomb and it will cause the explosion;
instead if it passes on the upper branch, the photon has a 50$\%$ probability to be absorbed by detector A and 50$\%$ to be revealed by detector B.
So, if the photon is detected by detector B it is certain that the bomb is usable, otherwise we can't conclude nothing.
After the examination of all bombs we will identify about 25$\%$ of usable bombs and detonate about 50$\%$ of these.
If we repeat the measurement many times, we are able to discriminate at maximum the  $33,3\%$ of usable bombs.
In 1994 the experimental proof of this system was realized\cite{BOMB2}.

\subsection{N09 Protocol: the proposed scheme}
\begin{figure}
   \begin{center}
   \begin{tabular}{cc}
   \includegraphics[scale=0.43]{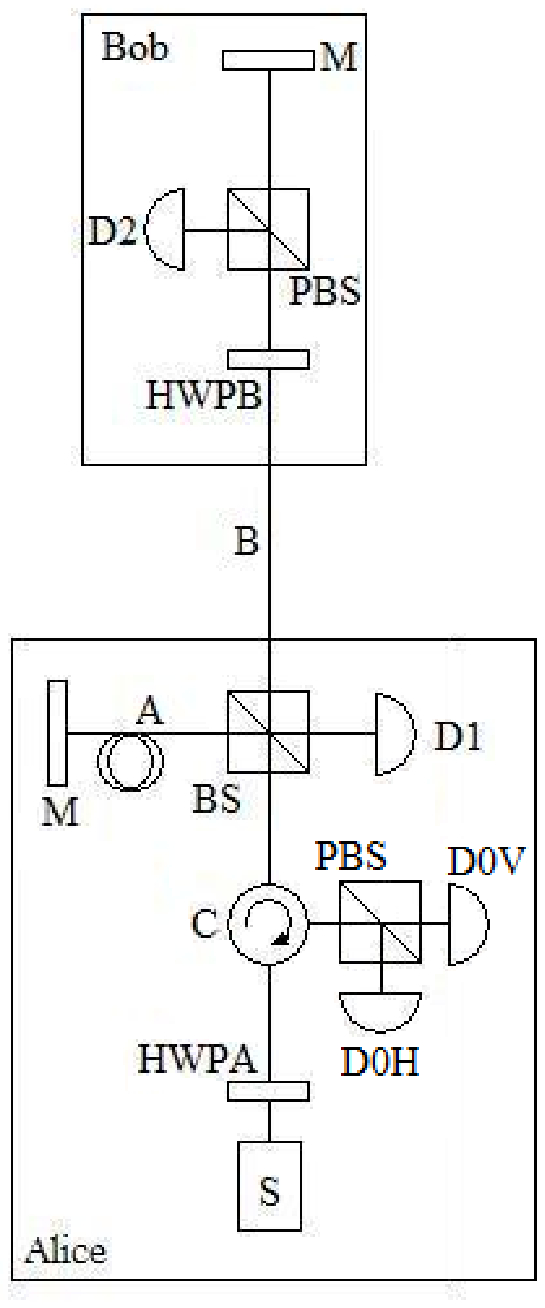}&
   \includegraphics[scale=0.2]{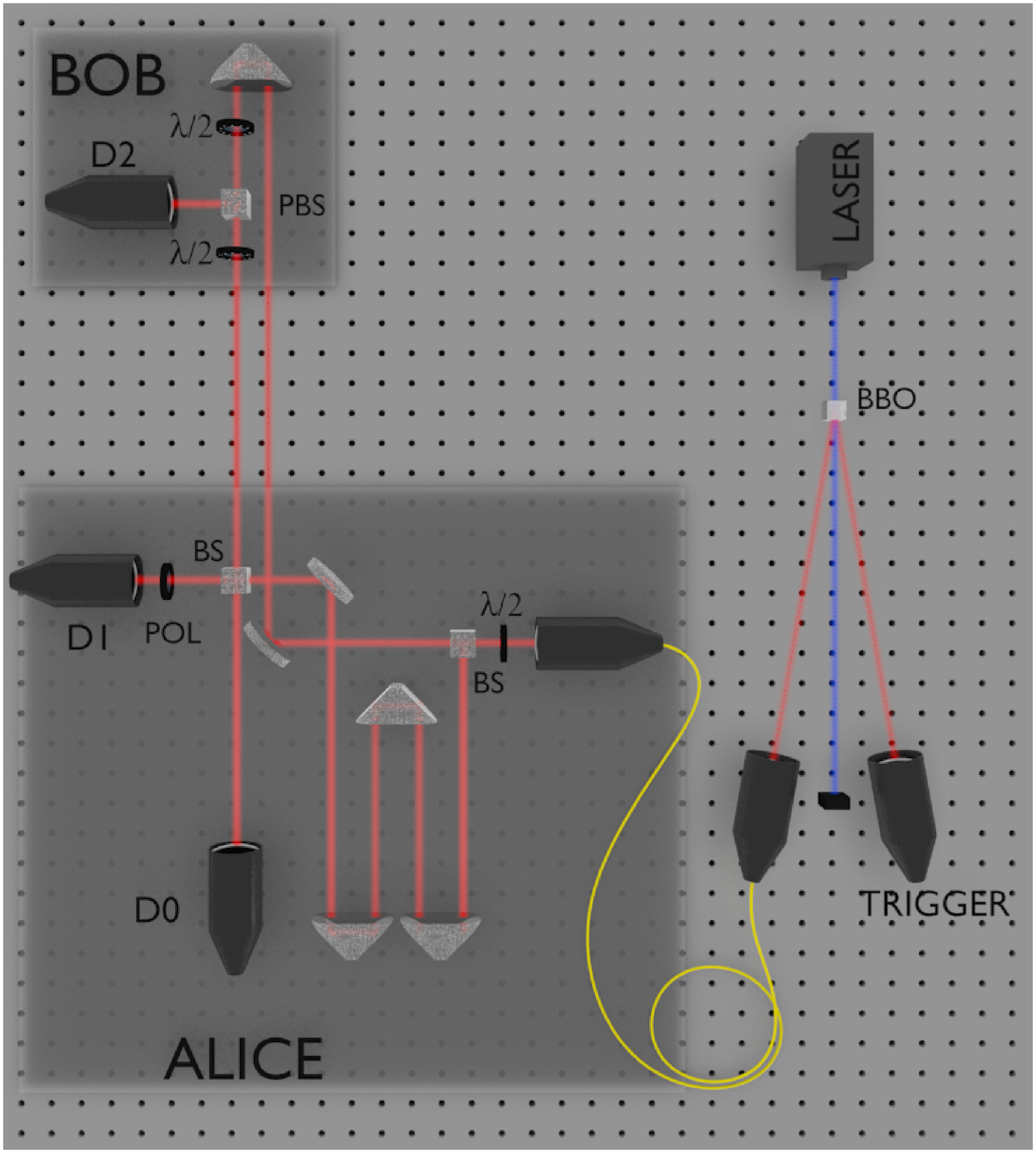}\\
   (a)& (b)
   \end{tabular}
   \end{center}
   \caption[example2]
   { \label{fig:setup}
   (a): scheme of the setup for Counterfactual QKD experiment equivalent to the one proposed by by Noh \cite{no}. (b) Setup of the implemented version of the protocol.A $100$ $mW$ laser source at $406$ $nm$ is used to pump a BBO type I crystal. In the degenerate regime, one of the PDC twin modes is used as heralded single photon and travels in the Mach-Zehnder interferometer including the two parties and the quantum channel. For each photon A and B perform randomly and independently either of two possible polarisation rotations ($0$ or $\pi/2$), by means of half-wave plates ($\lambda/2$). Detection events at the output ports of the interferometer and at the control output are revealed by Single Photon Avalanche Photodiodes (D0, D1, D2). The choice of equal polarization rotations leads to interference and consequently in the deterministic clicking of D0, while the choice of complementary angles results in the statistical distributions of the clicks among the three detectors. In the latter case, the events revealed by D1 correspond to the exchange of one bit of information between the users without the transmission of any photon through the channel.
   }
   \end{figure}

The protocol was proposed by Tae-Gon Noh in 2009 (N09) and it presents a main difference from the previously described protocols:
by implementing this protocol it is possible to perform the secret key distribution without letting the particle that carries the information even travel through the quantum channel.
This has the clear advantage from the point of view of security that  no eavesdropper can have direct access to the quantum system
of each particle involved in the communication.
In figure \ref{fig:setup} (a) the scheme proposed by Noh to realize his protocol is shown.

Before the description of the implemented protocol it is useful to give a brief description of the ideal scheme according to the setup originally proposed by the author (figure 1 of paper \cite{no}), which requires the use of a Michelson interferometer.


Alice randomly rotates the single photon
polarization (which originally is to be assumed horizontal) by means of a half wave plate (HWPA), either by $0$ (bit value
"0") or by $\pi/2$ (bit value "1").
Then, the photon enters one port of a $50:50$ beam splitter (BS), which is the first
element of a Michelson interferometer. After BS, according to
the polarization, the photon is in one of the two orthogonal states:
\bea | \phi_0 \rangle &= (| 0 \rangle_A | H \rangle_B + i |H \rangle_A | 0 \rangle_B ) / \sqrt{2}
\\ |\phi_1 \rangle &= (| 0 \rangle_A | V \rangle_B + i |V \rangle_A | 0 \rangle_B ) / \sqrt{2} \eea
 The path A of the interferometer (containing an optical delay OD and a mirror) is inside Alice's sector, while path B
 reaches Bob's one.




Bob then randomly selects one of the two polarisations and detects the photon in this polarisation allowing the photon
in the complementary polarisation to fly back to Alice's site. This is realized by the HWPB and the Polarizing beam splitter (PBS). As the PBS addresses the $\ket{V}$ photon towards D2, while \ket{H} photon is sent towards the mirror (M), rotations of the polarization of $0$ and $\pi/2$ induced by the HWPB correspond to the detection of $\ket{V}$ and $\ket{H}$ photon state by D2. If the photon is not detected by D2 but reflected back by M
it passes through the HWPB in the selected position, thus the photon gains back its original polarization state  interfering with itself at BS at Alice's site and, for a proper tuning of
the optical delay OD, it deterministically exits in D0.

When Alice and Bob select complementary polarization rotations, then either the photon is transmitted by
BS and
detected by Bob at D2 with 50\% probability (since its polarization at PBS is vertical), or it is reflected in path A.
In this case the photon, after passing through the OD twice, returns to the BS and then it is 
reflected or transmitted with equal probability (25\%). The first case leads to the clicking of D0, the second
corresponds to the photon arriving at D1H or D1V detectors depending on whether its polarization (selected via another PBS) is horizontal or vertical.

After the detection is completed Alice and Bob can communicate each other whether or not each of
the detectors clicked. If clicked either D0 or D2, with the purpose of detecting the intervention
of an eventual eavesdropper, they announce both the detected and the initial polarization state. If
D1H or D1V  clicks, Alice compares the initial and final polarization states: if they are consistent she does
not reveal any information, otherwise she announces her result. Alice and Bob can then establish a common key by using only the events when the photon was detected
at D1H or D1V (with the correct polarization).

The only apparent difference between the scheme discussed here and the original proposal in Ref \cite{no} is in the apparatus used by Bob to detect the photon at D2. Nonetheless the one shown accomplishes exactly the same task, thus
the two schemes should be considered absolutely equivalent. The original scheme also does not show explicitly the polarization selection system of the photons as we did with the introduction of D1H and D1V, but the polarization check is declared to be necessary.

 The very interesting point of this scheme is
 that the selection of events only at detector D1 correspond to
 photons that have traveled path A, i.e. never exited Alice's sector.
Therefore, the task of creating a secret key has been accomplished
without any photon carrying the information having been  outside
Alice's laboratory.

In Ref. \cite{sun} a more efficient and complicated CQKD was
proposed, whereas security issues of the N09 protocol were
considered in Ref. \cite{yin}, where it was proved its unconditional
security by considering its equivalence to an entanglement
distillation protocol. Finally, very recently, a security proof for
intercept-resend attacks in realistic situation (non unit detector
efficiency and presence of dark counts) was provided \cite{zhang}.

A first attempt to realize experimentally Noh's scheme is reported
in Ref. \cite{las}. However, this set up missed the key element of
CQKD, since the photon was indeed transmitted between Alice and Bob.

\subsection{N09 protocol: the experimental setup}
In the following we present the results of our equivalent implementation
of the protocol which is completely analogous to the one of Fig. 1(a), but it is based on a Mach-Zehnder
interferometer instead of
a Michelson interferometer and which is the first real proof-of-principle implementation of counterfactual QKD.

 In our experimental set up, as shown in Fig. 1(b), a heralded single photon source
exploiting parametric down-conversion (PDC) is used
: a
100 mW laser emitting at 406 nm in continuous-wave regime pumps a type-I BBO 
crystal producing
degenerate PDC at 812 nm. The emission of the PDC photons is slightly non-collinear
 corresponding to an emission angle of approximately $3^\circ$ with respect to the pump direction.
 The heralding photon after passing through
a 10 nm bandwidth interferential filter and a 4 mm wide pinhole is
coupled to a multi-mode fiber and addressed to the trigger detector.
The heralded photon, to be used as our true single photon state, is
selected by an interferential filter (1 nm FWHM
) and coupled to a single mode fiber leading to the
input of the interferometer.

The latter is a balanced Mach-Zehnder Interferometer (MZI) in which
each arm has an adjustable trombone prism. One of the two arms is
entirely included in Alice's site, while the other contains both the
quantum channel and Bob's site, the latter being composed by a PBS
between two half-wave plates (HWPB1, HWPB2) and D2 detector.

The choice of the Mach-Zehnder in place of the Michelson interferometer allows to simplify the apparatus, for example
the optical circulator in no more necessary even if the scheme is in principle equivalent to the proposed one.

The balance of the interferometer is guaranteed by a
closed--loop piezo--electric movement system, which stabilizes the
position of one of the trombones regulating the length difference between the two optical paths
inside the MZI with nanometric resolution.

The outputs of the interferometer, after spatial selection via $1$ $mm$ diameter-wide irises, are then coupled in multi-mode
fiber with no further spectral selection. A polarizer (pol) is also placed before D1 to check the polarization of the incoming photons.
 All the
signals (including the heralding photons and D2 clicks) are
revealed by Single Photon Avalanche Detectors (SPADs) with a
$\approx 60 \%$ detection efficiency at 812 nm.

Coincidence and time-tag analysis of the incoming signals are performed by means of PicoQuant
HydraHarp 400 multichannel picosecond event timer.
All the reported data were acquired in measurements of 20 seconds.
Our results show good agreement with the theoretical predictions and represent a proof of principle of
the experimental feasibility of CQKD.

\begin{figure}
   \begin{center}
   \begin{tabular}{c}
   \includegraphics[scale=0.45]{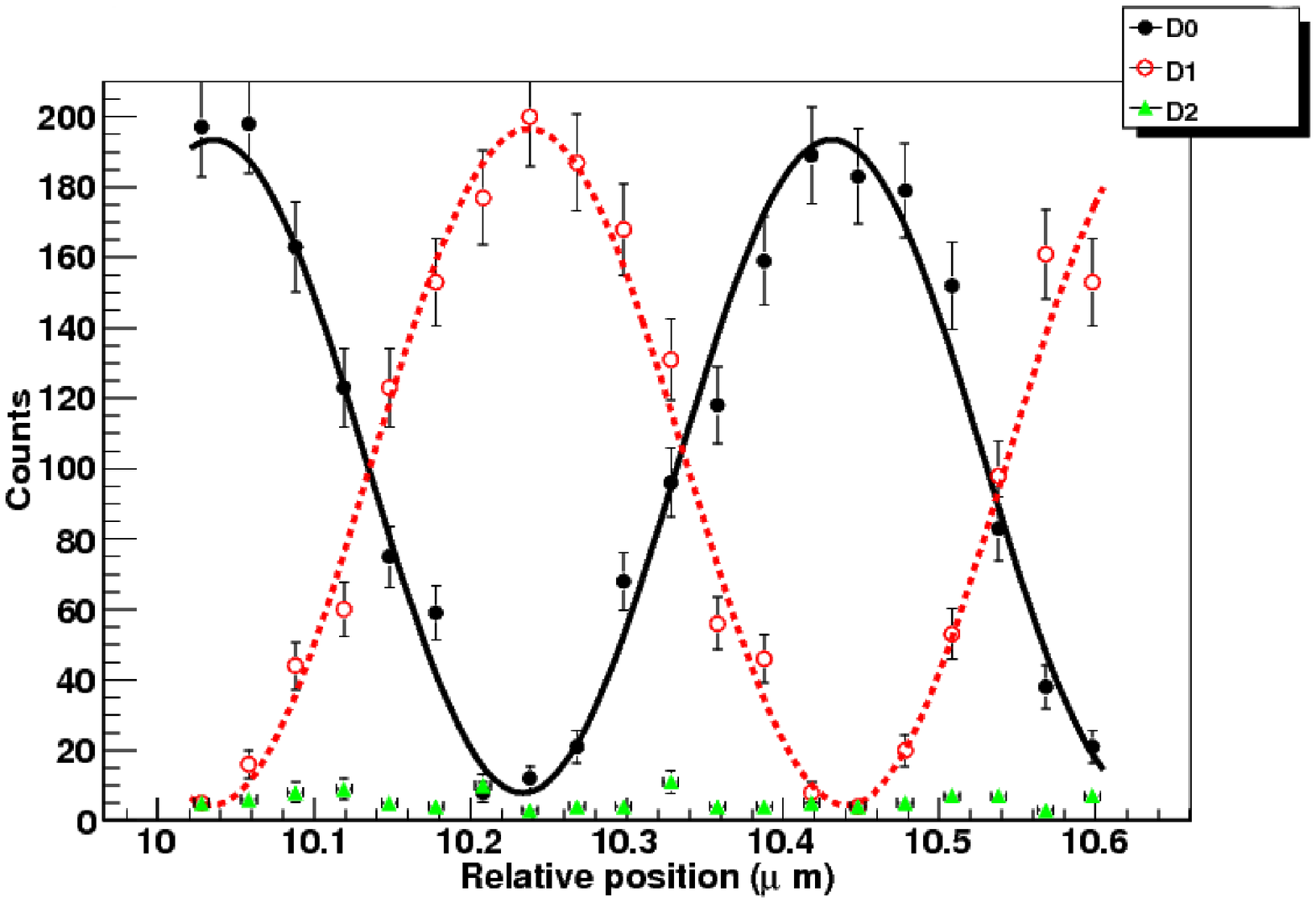}\\
   \includegraphics[scale=0.45]{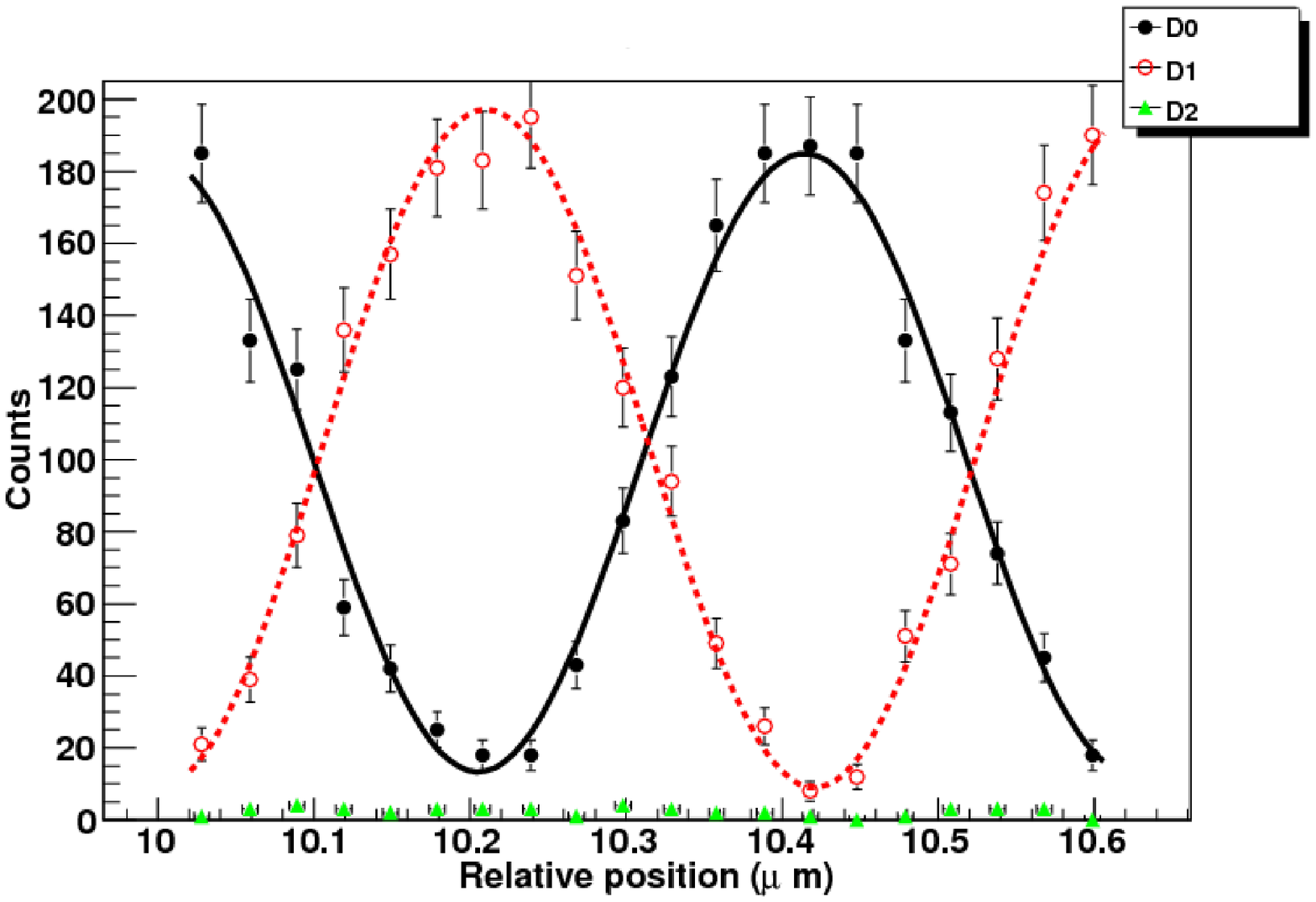}
   \end{tabular}
   \end{center}
   \caption[example2]
   { \label{fig:scan} Coincidence counts between the heralding channel and each of the MZI output detectors D0, D1 and D2 in 20 seconds acquisitions
as a function of the displacement of the prism balancing the interferometer when Alice and Bob
  use compatible sets of angles (top figure: $\{0,0\}$; bottom figure: $\{\pi/2,\pi/2\}$). For this choice of angles an interference pattern (with visibilities generally above $90\%$) can be observed in the
D0 and D1 counts and also control counts (D2) are  consistent with zero as expected.}
   \end{figure}

\begin{figure}
   \begin{center}
   \begin{tabular}{c}
   \includegraphics[scale=0.45]{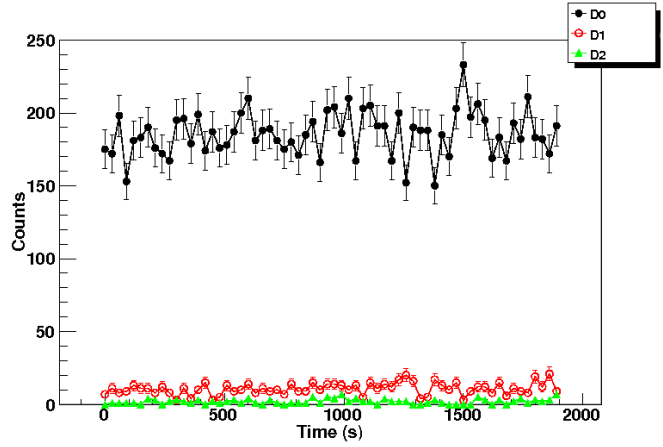}
   \end{tabular}
   \end{center}
   \caption[example2]
   { \label{fig:stab} Counting events showing the stability of the
interferometer in a half-an-hour long measurement when the balance of the two optical
paths is fixed.}
   \end{figure}

\begin{figure}
   \begin{center}
   \begin{tabular}{c}
  \includegraphics[scale=0.45]{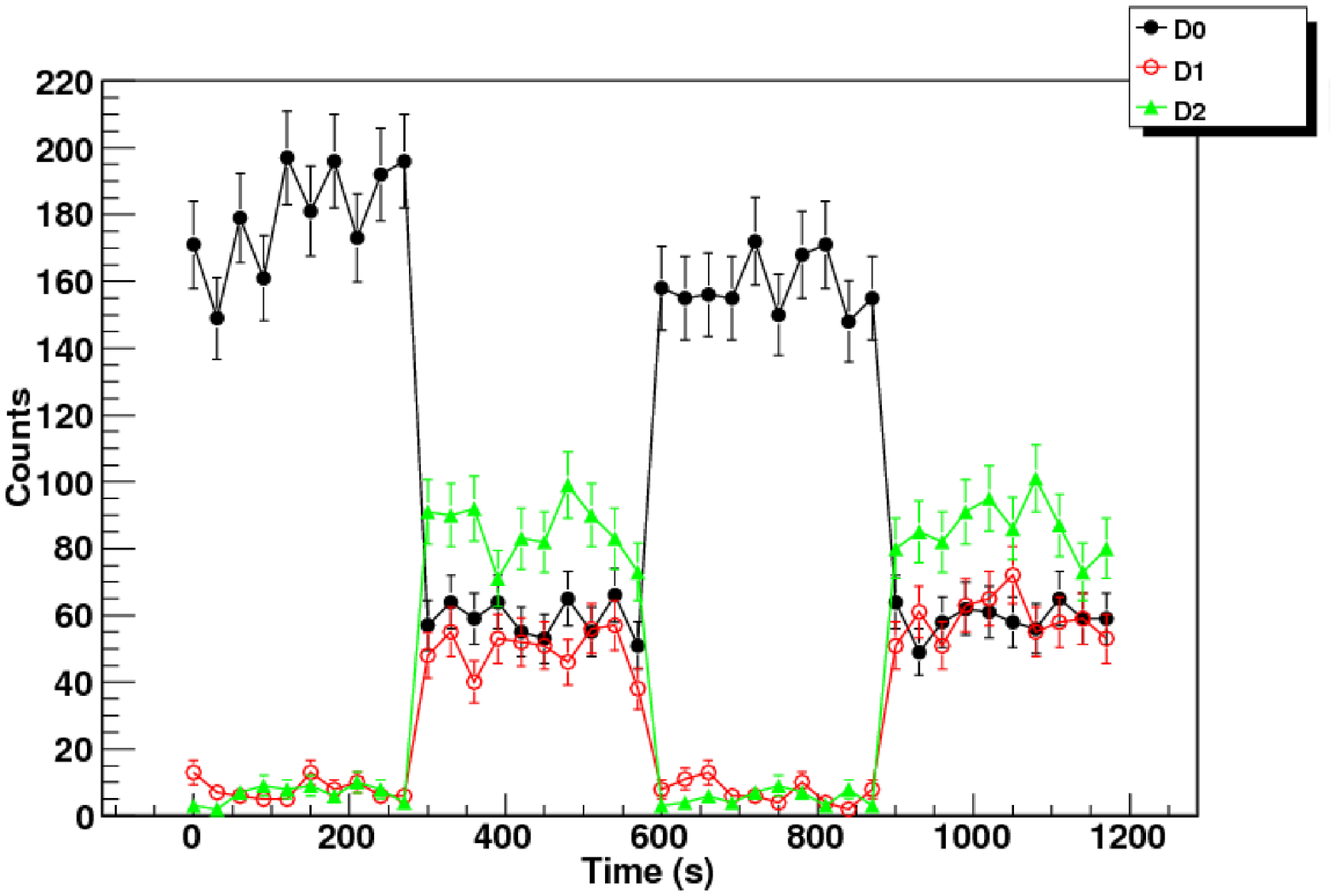}
   \end{tabular}
   \end{center}
   \caption[example2]
   { \label{fig:tras} Proof of transmission for the four possible angle choices for A and B. The segments
in which D0 and D1 counts are approximately equal, corresponding to the angles ([$0$,$\pi/2$];[$\pi/2$,$0$]), are the ones relative
to the actual transmission of information. In those events, the clicking of D1 delivers
a bit of shared information between the users even if no real photon photon travels in the
quantum channel. The estimated value for the QBER of the communication is ($12 \pm 1$)\%}
   \end{figure}

In Fig. 2 interference fringes with high visibility 
can be observed in the
coincidence counts between the heralding channel and each of the MZI output detectors D0 and D1
as a function of the displacement of the prism balancing the interferometer (within the coherence length of the signal,
which, according to the filters used, is of the order of hundreds of $\mu$m) when Alice and Bob
  use compatible sets of polarization rotation angles ($\{\theta_A,\theta_B\}=\{0,0\}$ or $\{\theta_A,\theta_B\}=\{\pi/2,\pi/2\}$). It can also be noticed that for this choice of angles the
D2 counts are consistent with zero as expected. In particular, when no rotation at all is performed ($\{0,0\}$),
the maximum visibilities are $(92 \pm 4)\%$ for D0 and $(96 \pm 4)\%$ for D1, while interference gets slightly spoiled 
for $\{\pi/2,\pi/2\}$ where the visibilities for D0 and D1 are respectively $(87 \pm 4)\%$ and $(91 \pm 4)\%$, values which, nonetheless, are sufficient for the proof of the protocol.
The uncertainty on the visibilities is obtained 
assuming a Poissonian distribution  for
the coincidence counts
.
Fig. 3 shows the stability of the
interferometer in a half-an-hour long measurement when the balance of the two optical
paths is fixed.

The performances of our key distribution process are summarized
in Table I.
Corresponding to the angles $\{0,\pi/2\}$ and $\{\pi/2,0\}$, D1 and D0 counts are approximately equal, 
as in this condition no interference should be present.
These are the events relative
to the actual transmission of information. In fact, the clicking of D1 delivers
a bit of the secret key  between the users even if no real photon travels in the
quantum channel.

In order to characterize the communication it is necessary to estimate the Quantum Bit Error Rate (QBER)
defined as the ratio between the probability for Bob to register an incorrect bit and the sum of the probabilities
of getting either a correct or an incorrect bit. In our case Bob gets an incorrect bit when D1 clicks even if Alice and Bob
use the same angle of polarization rotations and the events related to the correct transmission are those
in which D1 clicks when interference is destroyed.
Furthermore, we notice that when Alice and Bob use complementary polarizations the amount of photons with the wrong polarization detected by D1 is effectively null when dark counts are subtracted.
We can thus define QBER as

\begin{equation}
QBER=\frac{P_{D_1,int}}{P_{D_1,int}+P_{D_1,nint}}
\end{equation}
where $P_{D_1,int}$ is the probability for D1 to register a photon when Alice's and Bob's polarization rotations are equal, such that there is (destructive) interference, and $P_{D_1,nint}$ is the analogous probability in the case in which
Alice and Bob choose different angles.

For our measurements the mean QBER is  $QBER=(12 \pm 1)\%$.
We underline that all the reported measurements are obtained without subtraction of
background and accidental counts. If we account for these contributions, the corrected
QBER value decreases noticeably to $QBER'=(7 \pm 1)\%$, 
as would be the case if more reliable detectors were used,
 such as detectors affected by a lower dark count rate
. As already mentioned, the protocol has been demonstrated absolutely secure when ideal single photon sources are employed. To address the security problems eventually raised by the practical implementation of the protocol, firstly we tested it against
possible photon-number-splitting attacks, 
i. e. we investigated the quality of of our heralded single photon source.
From the measured count rates we obtained a value of 
$g_2(0)=(7 \pm 5)*10^{-9}$, which clearly shows negligible presence of multi-photon components.
The reason for such a small value is
related to the very low level of count rates (180 maximum in 20 seconds acquisitions) at the detectors.
This is basically due to the poor coupling efficiency of the heralded source (approximately $5\%$),
 the strict spectral selection on the heralding photons (1 nm FWHM filtering with $26\%$ transmittance),
 and also because of the spatial selection at the interferometers output (we used irises as narrow as 1mm in diameter to optimize
 the visibility of the interference fringes).
Furthermore, a small temporal detection window (1 ns) was selected in correspondence  of the arrival of the heralding photon.
Because of this temporal post-selection
 we mention that unheralded photons may travel inside the channel and Eve may exploit that to get 
significant information by intercepting them.
In order to overcome this security issue, shuttered heralded single-photon sources \cite{shuttered} should be considered a valuable solution, as they present comparable performances with respect to the non-shuttered ones.
Future developments of the scheme will include shuttered sources  together with 
stabilized fiber interferometers for wider distance.

We also address the issue of robustness of the protocol against more general
attacks
by computing the difference $m=I_{AB}-I_{AE}$, where $I_{AB}$ ($I_{AE}$) is the mutual information
between Alice and Bob (Alice and Eve), in the cases of general Intercept-Resend attacks and "Time-Shift" attacks.
Following the models suggested in Ref. \cite{zhang}, one can express $m$ for the intercept-resend attack as
\be
m_{IR}=P_{D1}[1-h(\frac{P_{e1}}{P_{D1}})],
\ee
where $P_{D1}$, $P_{e1}$ are respectively the click probability and the error probability at D1 and $h(x)
$ is the binary Shannon Entropy. We mention that $P_{e1}$ in principle includes not only the probability to register counts at D1 when they are not expected (${0,0}$,${\pi/2,\pi/2}$), but also the probability to detect a photon at D1
 with the wrong polarization. The latter probability has been estimated by counting the rate of photons impinging in D1 when the polarizer before it is set orthogonally to the polarization selected by Alice for each of the four angle combinations. According to the measured counts ($1.45 \pm 1.16$, $1.2 \pm 1.5$, $0.95 \pm 1.24$, $0.8 \pm 1.2$; in 20 seconds acquisitions), we estimated a mean value for this probability of $(4 \pm 3)*10^{-6}$.

Regarding the time-shift attack, where Eve exploits the non-ideality of the detectors,
one must subtract from the previous value two contributions, 
obtaining:
\be
m_{TS}=m_{IR}-\gamma-\Delta I_{AE}(\eta),
\ee
where $\gamma$ accounts for the maximum corrupted bit rate due to dark counts and $\Delta I_{AE}(\eta)=\frac{1-\eta}{2\eta}(P_{D2}-P_{e2})$  is the increment of the mutual information
between A and E due to non-unit efficiency of the detectors.

Both values 
calculated from the collected data are positive ($m_{IR}=0.23 \pm 0.04$, $m_{TS}=0.15 \pm 0.06$), 
ensuring 
the possibility of distributing a secret key \cite{gis,scar}

Altogether our results provide a satisfying proof-of-principle of the QKD scheme realized
in free-space
.
Nonetheless, recent results on the implementation of high stability fiber based Mach-Zehnder interferometers
 (over distances of the order of some km) \cite{mzfibre1, mzfibre2} 
 certify the possibility of exploiting this protocol 
in "real-life" (as well as commercial) applications.


\begin{table}[htbp]
\centering
   \begin{tabular}{|c|cccc|}
   \hline
    & $\{0,0\}$ & $\{0,\pi/2\}$ & $\{\pi/2,\pi/2\}$ & $\{\pi/2,0\}$\\
   \hline
   $C_{D0}$ & $180 \pm 4$ & $59 \pm 2$ & $159 \pm 4$ & $59 \pm 2$\\
   \hline
   $C_{D1}$ & $7.9 \pm 0.9$ & $53 \pm 2$ & $7.2 \pm 0.9$ & $59 \pm 2$\\
   \hline
   $C_{D2}$ & $6.6 \pm 0.8$ & $85 \pm 3$ & $5.4 \pm 0.7$ & $86 \pm 3$\\
   \hline
   $V_{D0}$ & $(92 \pm 4)\%$ & $(0 \pm 4)\%$ & $(0 \pm 4)\%$ & $(87 \pm 4)\%$\\
   \hline
   $V_{D1}$ & $(96 \pm 4)\%$ & $(0 \pm 4)\%$ & $(0 \pm 4)\%$ & $(91 \pm 4)\%$\\
   \hline
   \hline
   $QBER$ & & & \hspace{-12mm}$(12 \pm 1)\%$ & \\
   \hline
   \end{tabular}
\caption{Resume of the main results in the implementation of the CQKD protocol proposed in Ref. \cite{no}.
Each column refers to a set $\{\theta_A,\theta_B\}$ of polarization rotation performed by the users and  $C_{Di}$
labels the mean coincidence counts at the $i$-th detector in acquisition of 20 seconds. $V_{D0}$, $V_{D1}$ are the visibilities of the interference fringes observed at the two outputs of the interferometer by scanning the path length difference between the two arms of the MZI. $QBER$ is the estimated quantum bit error rate
for the transmission. }
   \end{table}

\section{Conclusion}
In this paper we presented an overview on our recent results regarding the first proofs of principle of two novel QKD schemes using only orthogonal states. The experimental results demonstrate the security of those protocols while, on the one hand, they prompt to a further optimization of the schemes, on the other hand they offer groundbreaking contribution in the discussion on the resources actually needed to perform secure QKD.
The research leading to these results has received funding from the European Union on the basis of Decision No.
912/2009/EC (project IND06-MIQC), by MIUR-FIRBRBFR10UAUV, and by Compagnia di San Paolo.

\end{document}